# Atom probe tomography:
# a local probe for chemical bonds in solids

Oana Cojocaru-Mirédin[1,2,*], Yuan Yu[1], Jan Köttgen[1], Tanmoy Ghosh[1,3], Carl-Friedrich Schön[1], Shuai Han[4], Chongjian Zhou[4], Matthias Wuttig[1,5,*]

[1] I. Physikalisches Institut IA; RWTH Aachen University, Sommerfeldstraße 14, 52074 Aachen, Germany

[2] INATECH, University of Freiburg, Georges-Köhler Allee 102, 79110 Freiburg, Germany

[3] Department of Sciences and Humanities, Rajiv Gandhi Institute of Petroleum Technology (RGIPT), Jais, Amethi - 229304, UP, India

[4] State Key Laboratory of Solidification Processing, Northwestern Polytechnical University, Xi'an , 710072, China

[5] Peter Grünberg Institute - JARA-Institute Energy Efficient Information Technology (PGI-10), Forschungszentrum Jülich GmbH, Wilhelm-Johnen-Straße, 52428 Jülich, Germany

**\*Corresponding authors**: cojocaru-miredin@physik.rwth-aachen.de; wuttig@physik.rwth-aachen.de

## Abstract

Atom probe tomography is frequently employed to characterize the elemental distribution in solids with atomic resolution. Here we review and discuss the potential of this technique to locally probe chemical bonds. Two processes characterize the bond rupture in laser-assisted field emission, the probability of molecular ions, i.e. the probability that molecular ions (PMI) are evaporated instead of single (atomic) ions, and the probability of multiple events, i.e. the correlated field-evaporation of more than a single fragment (PME) upon laser- or voltage pulse excitation. Here we demonstrate that one can clearly distinguish solids with metallic, covalent, and metavalent bonds based on their bond rupture, i.e. their PME and PMI values. Differences in the field penetration depth can largely explain these differences in bond breaking. These findings open new avenues in understanding and designing advanced materials, since they allow a quantification of bonds in solids on a nanometer scale, as will be shown for several examples. These possibilities would even justify calling the present approach *bonding probe tomography* (BPT).

## 1. Introduction

Atom probe tomography (APT) is a well-established nano-analytical technique enabling the determination of the spatial distribution of atoms in a solid with Angstrom resolution. It can characterize a broad spectrum of materials, ranging from metals to biological materials in three-dimensions[1,2,3,4,5,6]. Interesting physical phenomena such as impurity segregation[7,8], solute clustering[9,10,11,12,13], diffusion[14,15], and intermixing at hetero-interfaces [16,17,18,19], etc. can be evaluated thanks to its unique capabilities. Some of these studies even led to the discovery of phenomena such as the snowplow effect[20], strain-induced asymmetric line segregation[21], linear complexions[22], and Janus nano-precipitation[23].

In this review, we summarize the present understanding of the role of chemical bonds on laser-assisted bond rupture in atom probe tomography. It will be demonstrated that APT is suitable to probe the bond rupture and hence chemical bonds in solids on the nanometer scale. The high electric field of approximately 10 V/nm in conjunction with the pulsed femtosecond laser applied on the apex of a very small needle of ~60 nm in diameter allows for the so-called laser-assisted field evaporation[24]. Upon laser-assisted field evaporation, atoms at the surface of the needle's apex are dislodged by breaking bonds to their neighbors. Hence, with APT well-characterized bond rupture experiments are conducted, potentially justifying the acronym (BPT), i.e. bonding probe tomography. We are not seriously suggesting to use the acronym BPT, but want to stress in this review that APT has tremendous potential to probe chemical bonds on a very local scale. Here we want to demonstrate that this is the case and provide arguments why this is so. Finally, it will be shown how APT can be employed to provide crucial insights to understand and tailor materials.

Studies on bond strength and bond-breaking behavior with APT have so far been performed on ceramics[25], biological materials[26] and phase change materials[27,28]. However, only for phase change materials (PCMs), which can also be used as excellent thermoelectric compounds[29], a very unusual bond rupture has been observed that will be described in detail next. In the subsequent section, the unusual bond rupture will be related to an unconventional type of bonding. There, it will be shown that the bond rupture in atom probe tomography differs significantly for solids which employ metallic, covalent and metavalent bonds. In conjunction with the high spatial resolution of atom probe tomography, this allows the bonding mechanism to be determined at the local level, which would justify the abbreviation BPT. In section 4, a pertinent question will be answered: How can these differences in bond rupture for the different bonding mechanisms be explained? In section 5, we will finally discuss how these differences can be utilized to understand and design advanced functional materials.

## 2. Quantities characterizing the bond rupture in atom probe tomography

The principle of APT is based on the field evaporation of charged atoms (ions) from the surface under a high electric field of about $10^{10}$ V/m[24]. Under this high electric field, the atoms on the surface are restrained in a partial ionic state with a reduced energy barrier. In order for field evaporation to take place, the ions must overcome this energy barrier, which is facilitated by a voltage or a laser pulse. It is well-accepted that not only atomic ions but also molecular ions are field-evaporated and registered by a position-sensitive detector during an atom probe experiment[30]. Interestingly, the probability that fragments are dislodged from the tip as molecular ions is material specific. Thus, the probability that molecules rather than atoms are released during laser-assisted field evaporation is one characteristic of bond breakage. Subsequently, we will abbreviate this probability as the PMI (Probability of Molecular Ions, schematically depicted in Figure 1a). Metals are typically field evaporated as single ions, thus exhibiting very low PMI values. Yet, field evaporation from covalent or ionic materials often leads to the emission of molecular ions (see Figure 1b), i.e. a high PMI. Typical covalent or ionic materials with such high PMI values are oxides, nitrides, and carbides[31,32,33]. Interestingly, many chalcogenides including selenides, tellurides, and sulfides exhibit a high PMI value, too[6,34,35,36].

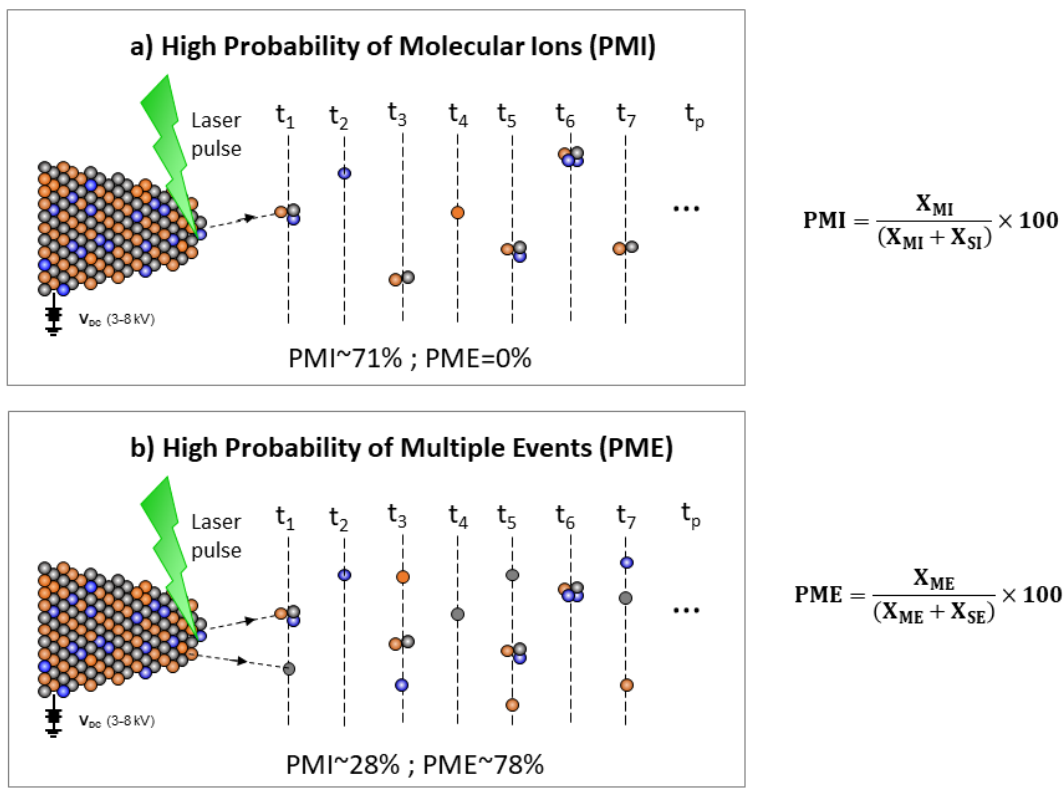

**Figure 1. Graphic presentation of the "Probability of Molecular Ions" (PMI) and the Probability of Multiple Events" (PME).** Sketch of a bond-breaking experiment for a sequence of successful pulses (pulses for which one or several events took place) exhibiting a) high PMI as well as b) high PME. SI stands for an ion which consists of a single ion, while SE stands for a single event (fragment) to be dislodged upon bond rupture.

Such molecular ions can dissociate on their flight path to the detector. Hence, it is possible that for the same laser or voltage pulse more than one ion reaches the detector, even though only a single molecular ion left the tip. The dissociation of molecular ions is based on the phenomenon of bond softening, which occurs in a molecular ion in a high electric field. This process has been studied thoroughly in the 1970s[37]. More specifically, in the absence of an electric field the charge of the binding electron is distributed symmetrically around the ion cores. Yet, this situation changes in the vicinity of a positively charged APT tip surface, where the molecular ion AB$^+$ is found in a high electric field (see Figure 2a for the case for $H_2^+$). Then, the probability of finding the electron adjacent to the tip surface is much larger than near the ion core B (i.e. a polarization effect occurs). This polarized molecule vibrates with a certain frequency (e.g. $10^{13}$ s$^{-1}$ for a $H_2^+$ molecule[37]).

During a vibration work has to be done against the intramolecular forces. However, now the potential energy is lowered due to the high electric field (Figure 2b). This effect is easily quantified if the two ion cores A and B have the same mass. Then, the potential energy is lowered by $-\frac{1}{2}eF(r-r_0)$, where $e$ is the elementary charge (C), $F$ is the field (V/nm), $r$ is the distance between ions A and B and $r_0$ is the equilibrium distance of ions A and B as displayed in Figure 2b. Thus, for a sufficiently high field strength, the maximum of the potential curve vanishes, i.e. the dissociation energy $E_b'$ becomes 0, and the dissociation of the molecular ion takes place immediately.

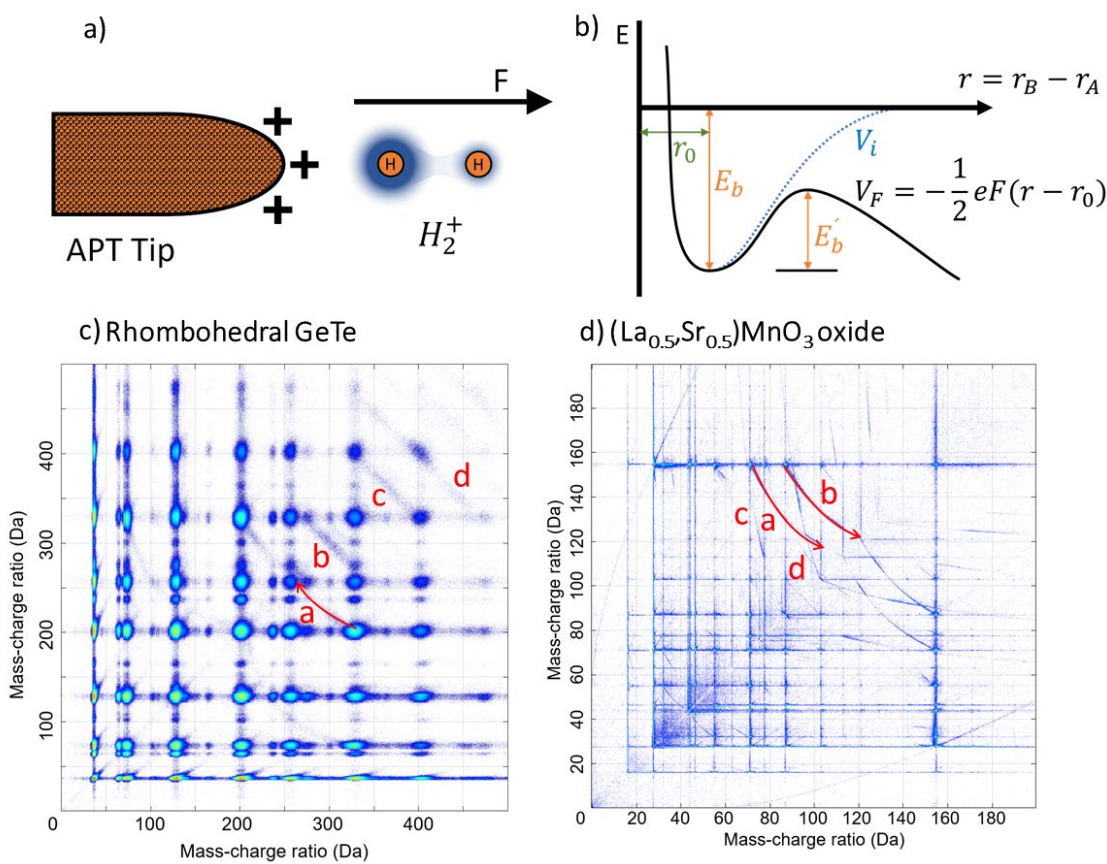

**Figure 2. Field dissociation of molecular ions.** a) Schematic presentation of the dissociation mechanism of an $H_2^+$ ion in a high electric field. b) Representation of the potential curve of a molecular ion consisting of two atoms of equal masses. Superposition of the potential $V_i$ of the polarized molecular ion as shown in a) and of potential $V_F$ of the molecular ion under the external field. Hence, $E_b$ is the dissociation energy in the absence of a high field, while $E_b'$ is the dissociation energy in the presence of a high field. Ion correlation histogram of c) rhombohedral GeTe showing the weak dissociation tracks such as the dissociation a) $Ge_2Te_3^{2+}$ → $GeTe_2^+ + GeTe^+$ and b) $Ge_3Te_3^{2+}$ → $GeTe_2^+ + Ge_2Te^+$ as well as of d) $(La_{0.5},Sr_{0.5})MnO_3$ oxide showing strong dissociation tracks such as the dissociation a) $LaMnO_2^{2+}$ → $LaO^+$ + $MnO^+$ and b) $LaMnO_3^{2+}$ → $LaO^+$ + $MnO_2^+$. Figures in a) and b) were adapted from Ref.[37].

A very elegant way of observing such molecular dissociation processes has recently been proposed by utilizing the ion correlation histogram[38,39]. Examples of ion correlation histograms are given in Figure 2c,d for laser-assisted field evaporation of rhombohedral GeTe and $(La_{0.5},Sr_{0.5})MnO_3$. Here combinations of ions $m_1$ and $m_2$ are used to construct these 2D histograms. The opposite ordering of each pair is also considered resulting in a histogram which is symmetric with respect to the diagonal axis, i.e. $m_1 = m_2$. The dissociation of molecular ions leads to tracks emanating from this diagonal axis to the ion-pair coincident point ($m_1$, $m_2$). In this process, a larger ionic fragment (parent ion) with a given kinetic Energy ($E_{kin}$) breaks apart into two smaller fragments (daughter ions) under the high electric field. As shown in detail in Ref.[38] a daughter fragment with a smaller mass-to-charge ratio will arrive later than expected on the detector due to the relatively slower flight speed compared with the parent ion. On the contrary, the other fragment with a higher mass-to-charge ratio will arrive earlier than expected. The result is that the ion pair moves away from the diagonal forming characteristic tracks as seen in the dissociation histogram. In Figure 2c for example very weak dissociation tracks are visible for the GeTe compound (most visible are a) $Ge_2Te_3^{2+}$ → $GeTe_2^+$ + $GeTe^+$ and b) $Ge_3Te_3^{2+}$ → $GeTe_2^+$ + $Ge_2Te^+$), while on the contrary strong dissociation tracks are visible for $(La_{0.5},Sr_{0.5})MnO_3$.

Such molecular dissociation processes also lead to an increase in the number of multiple events. This probability of multiple events (PME) describes the probability that a laser (or voltage) pulse creates more than a single fragment registered on the detector. Having a low probability of multiple events–i.e. having a high probability that only a single ion is detected per successful laser pulse is usually considered to be the signature of a high-quality APT dataset. This is because a high PME may lead to "incorrect" positional information of an ion[40] and/or compositional inaccuracies[41,42,43]. There are various possible mechanisms or even artefacts that cause high PME values, as discussed in the literature (see Table 1). The earliest one studied is the molecular ion dissociation mechanism presented above. Here molecular ions dislodged from the tip dissociate on the flight path to the detector, leading to a high PME. As mentioned above, this mechanism leading to high PMEs can be separated from other mechanisms leading to a high PME due to its distinct correlation histogram. Molecular dissociation is typically observed in nitrides, oxides, and other materials prone to field evaporation, producing molecular fragments such as GaSb[31,32,33,44]. Several other effects which lead to the detection of multiple events are in fact typical artefacts observed in APT. These processes include atomic migration on the apex of the tip prior to evaporation (observed for example for C atoms in Fe-C alloys[45]) and pile-up effects (as seen for example in carbides[46] and borides[42]). For some of the compounds, the multiple events were found to be correlated in space and time for most of the compounds given in Table 2, leading to inaccuracies in the overall atom probe composition determination[41,42,43]. Spatial correlation describes the probability of finding a certain distance of ions identified by the 2D detector, which were created by a single (laser or voltage) pulse. Temporal correlation instead describes the probability that there is a certain number of pulses $n_p$ between two successful pulses (pulses for which at least one ion is detected on the detector) for which no ion evaporation took place (called null pulses). If the multiple events are correlated in time, then the number $n_p$ decreases strongly so that the evaporation of multiple events for one successful pulse leads to the evaporation of other multiple events within the next successive pulses.

| Sample | | PME (%) | Mechanism for multiple events formation | Reference(s) |
|---|---|---|---|---|
| **Nitrides** | Ti-Si-N | 59 | Molecular Dissociation | 41 |
| | AlGaN | 27-57 | - | 47 |
| | GaN | 40 | Molecular Dissociation | 40,38,48 |
| **Oxides** | $Y_1Ba_2Cu_3O_{7-\delta}$ | 28-40 | Molecular Dissociation | 40 |
| | ZnO | 40 | Molecular Dissociation | 31,49 |
| | $Cr_2O_3/TiO_2$ | >45 | Molecular Dissociation | 50 |
| | $Al_2O_3$ | >50 | Molecular Dissociation | 51 |
| **Others** | GaSb | 6-45 | Molecular Dissociation | 43 |
| | NiSiPt | 14-42 | Unclear; probably differences of evaporation field | 52 |
| **Carbides** | Fe-C alloys | - | Artefact: C surface migration prior to field evaporation | 45 |
| | WC | 52-60 | Artefact: detector pile-up and dead time (the time spread between two ions of the same | 46 |
| | Ti(C,N) | 55 | | |
| | $Ti_2AlC$ | 41 | | |

|         | SiC     | 45 | mass-to-charge ratio coming from the same pulse is smaller than the dead time of ~3 ns). |    |
|---------|---------|----|-------------------------------------------------------------------------------------------|----|
|         | $M_{23}C_6$ | 30 |                                                                                       |    |
| Borides | BN      | 80 | Artefact: pile-up effect or preferential field evaporation                                | 42 |
|         | Boron   | 64 |                                                                                           |    |

**Table 1:** Summary of all material classes (except metavalent solids) which show a higher PME and their corresponding PME values.

Surprisingly, there is one class of materials which behaves very differently. In this class of solids, high PME values of more than 50% are characteristic of an unconventional bond rupture[53]. In this class of materials which includes crystalline GeTe, $Sb_2Te_3$ and elemental Bi, a PME larger than 50% has been observed. Such a high value has also been observed for PbSe in recent work by Hughes et al.[54]. In total, more than 50 solids have been identified[29,55,56], for which the high PME values cannot be explained by the molecular dissociation mechanism or the artefacts described above. An example is crystalline $GeSe_{0.5}Te_{0.5}$, which reveals a PME value of 55%, but exhibits no molecular dissociation. This statement can be derived both from the correlation histogram (in Figure 2c,d) and the frequency diagram shown in Figure S1. Interestingly, for the rhombohedral $GeSe_{0.75}Te_{0.25}$ characterized by a very high PME value only one peak was observed in the blue curve, i.e. the one associated with correlated evaporation. Surprisingly, the peak intensity of the red curve for the rhombohedral $GeSe_{0.75}Te_{0.25}$ compound (Figure S1a) is almost 10 times lower than the peak intensity of the blue curve, proving that the multiple events are not strongly correlated in time (no burst evaporation). Even more astonishing is that the peak intensity of the red curve for the rhombohedral $GeSe_{0.75}Te_{0.25}$ compound (characterized by a very high PME valueof 58%) is 2 times lower than that of its amorphous counterparts (characterized by a low PME value), proving that the multiple events are weakly correlated in time despite their very high proportion. Yet, the very high peak intensity of the blue curve for both compounds (Figure S1a) suggests a strong spatial correlation between multiple events as expected.

Comparing the field evaporation behavior of rhombohedral $GeSe_{0.75}Te_{0.25}$ with one of the metals like Al (Figure S1b) reveals that correlated evaporation is the evaporation mechanism responsible for both systems. The only difference between Al and rhombohedral $GeSe_{0.75}Te_{0.25}$ (and extrapolated to metals and crystalline PCMs in general) is that the PME value for rhombohedral $GeSe_{0.75}Te_{0.25}$ and crystalline PCM in general (PME>55%) is well above than that measured for Al or other metals (PME<15%). Hence, this high PME value is attributed to an unconventional mechanism named 'enhanced-correlated field evaporation mechanism'[27]. It is an intrinsic property of crystalline PCMs[27]. Interestingly such enhanced-correlated field evaporation mechanism cannot be applied for the amorphous counterparts where low PME values (below 30%) were measured. This shows that the bond rupture in amorphous and crystalline phase change materials of the same stoichiometry differs significantly. This striking observation requires an explanation, which will be presented in section 5.

One can hence conclude that three different bond rupture scenarios have been found in APT as summarized in Figure 3 below. Solids like Al, Au, Ag, Cu, W and NiAl show a very low PMI and a low PME. This bond rupture scenario is apparently characteristic for metals. Covalent semiconductors like GaAs, GaSb or InSb, on the contrary, have a much higher PMI but a low PME. Hence, solids that employ metallic or covalent bonding can be distinguished in atom probe tomography by analyzing the bond rupture. Finally, a significant number of crystalline chalcogenides shows yet another characteristic bond rupture. These solids have unusually high PME values of above 50%, which can be attributed to enhanced correlative field evaporation. They also have PMI values much higher than metals. Crystalline solids like GeTe, $Sb_2Te_3$, PbTe or Bi are characterized by this bond rupture. These solids possess an unusual property portfolio which has been attributed to an unconventional bond type, named metavalent bonding[53,57]. As a consequence, the bond rupture in APT can distinguish between metallic, covalent, and metavalent bonds. What is even more striking, the bond rupture in materials like GeTe differs between the crystalline and the amorphous state. While crystalline $Ge_2Sb_2Te_5$ has a high PME of 65%, amorphous $Ge_2Sb_2Te_5$ shows a much lower PME of 22%[53]. This implies that the crystallization of amorphous $Ge_2Sb_2Te_5$ leads to a change of bonding from covalent to metavalent bonding. This change of bonding can explain the pronounced property changes that accompany crystallization in such chalcogenides[58], which is utilized in phase change materials for data storage[59]. To understand the origin of these differences in bond rupture, we need to look at the different properties that characterize the different bond types in solids.

## 3. Bonding classification and its relation to bond rupture (PMI and PME)

It has recently been shown that it is possible to distinguish different classes of chemical bonds in solids due to the different portfolios of properties these compounds possess[60,57]. Metals, for example, are characterized by a vanishing band gap, a large effective coordination number (ECoN), high values of the electrical conductivity ($\sigma > 5 \times 10^4$ S/cm), moderate Grüneisen parameters for transverse optical mode ($\gamma_{TO}$, a measure of the anharmonicity of the lattice), and vanishing values of the Born effective charge $Z^*$, which characterizes the bond polarizability[57]. Solids, which employ covalent bonding instead, are characterized by a band gap, a coordination number that usually follows the 8 – N rule, a much lower electrical conductivity if the crystals are not doped, and a $\gamma_{TO}$ which is usually close to 2. The crystalline chalcogenides and related compounds discussed here, which show an unusual bond rupture, are also characterized by a unique property portfolio. Typical physical properties such as the optical dielectric constant $\varepsilon_\infty$, $Z^*$, and $\gamma_{TO}$ are much larger for these solids than for metals and covalent solids. This peculiar property portfolio as well as the unusual bond-breaking behavior are indicative of this novel class of chemical bonds termed termed "metavalent bonding"[53,57]. It is striking that amorphous solids of the same chalcogenides neither show these properties nor this unconventional bond rupture, which is further evidence of significant changes in bonding upon crystallization[58].

This raises the question of how the differences in bond rupture observed for the different solids can be explained. Ideally, we would like to understand, if there is a single material property which governs or is at least closely related to the bond rupture. Previous work has shown that no single parameter can distinguish all different bonding mechanisms [57,60]. Instead, a portfolio of five different properties has been shown to enable a classification of different bonds. Nevertheless, the one quantity which has the highest predictive power concerning bond breaking in the atom probe is the electrical conductivity of the solids. Figure 3a shows that semiconductors with covalent bonding (in red) and metals (in blue) exhibit low PME values, as expected. Yet, only crystalline PCMs (in green) show exceptionally high PME values above 55%. On the contrary, amorphous

PCMs (covalently bonded compounds) like amorphous $In_3SbTe_2$, amorphous GeSe, amorphous $GeSe_{0.25}Te_{0.75}$ and $GeSe_{0.5}Te_{0.5}$ show PME values below 30%[27]. Hence, metavalent solids are characterized by a distinct bond rupture. It is striking that this unusual bond rupture is observed in the transition region between metallic bonding, where electrons are highly delocalized, and covalent bonding, where electrons are localized between the ion cores. This leads to a pronounced maximum in the transition region, i.e. a 'volcano'-shaped curve, confirming the notion that metavalent bonds are located in the competition zone between metallic and covalent bonds.

The second relevant parameter employed here to distinguish different solids is the PMI. This quantity is close to zero for metals. Only for solids with a conductivity below $5 \times 10^4$ S/cm, significant non-zero values of the PMI are observed. Again, a volcano-like curve is observed, where exceptionally high values of the PMI are found for metavalent solids, i.e. in the transition region between metallic and covalent solids. Yet, figure 3b also offers two other important insights. The highest PMI value displayed is found for $Sb_2S_3$, a covalent semiconductor where a PMI of 99.9% is found. For this compound, up to 42% of the ions are dislodged as $Sb_3S_4^+$ ions. This shows that the PMI alone cannot distinguish the different bonding mechanisms, we need both the PME and the PMI to distingiuish all three different bonding mechanisms, as demonstrated in Figure 3.c. Furthermore, visual inspection of Figure 3 also shows that the electrical conductivity is insufficient to distinguish different types of bond rupture. In the region around $10^2$ S/cm, both metavalent and covalent solids are found which differ in PME and PMI. This raises the question of what role electrical conductivity plays in the breaking of the bond, a topic that will be addressed next.

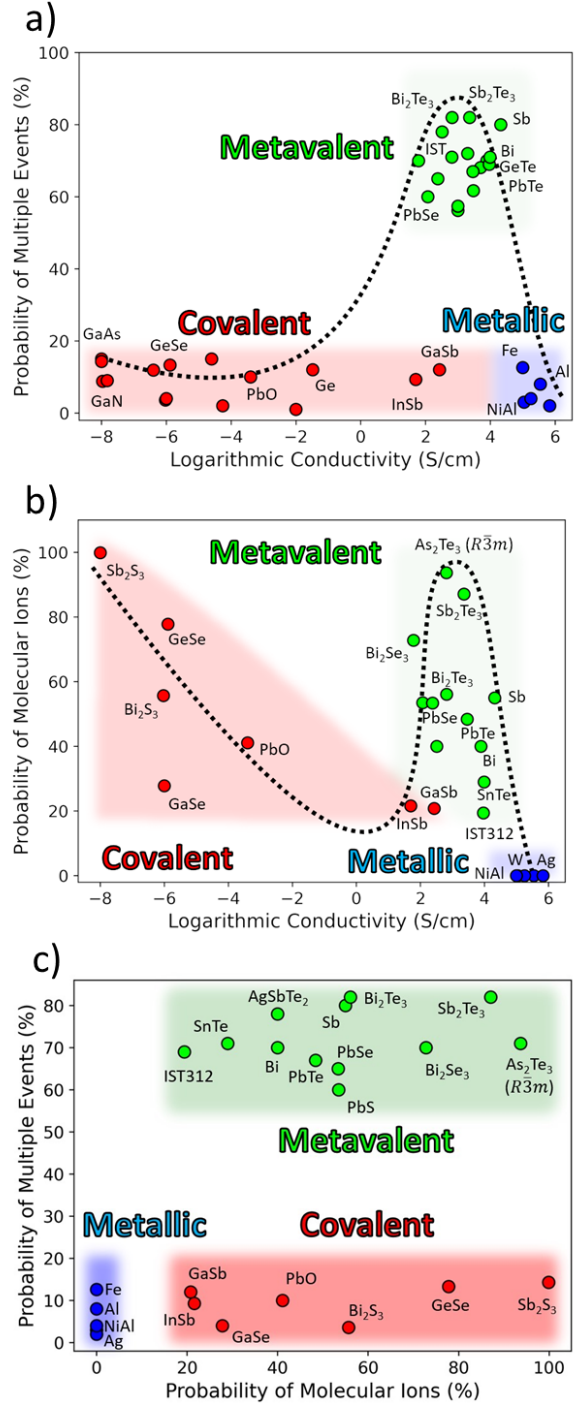

**Figure 3: Interdependence between the electrical conductivity of the solids and two quantities which characterize the bond rupture, a) the PME, b) the PMI, and c) PMI vs. PME.** Several classes of materials can be distinguished based on their bond breaking in the atom probe: metals (in blue), metavalent solids (in green), and covalently bonded compounds (in red). Please note that some of the amorphous phase change materials show a bond rupture which is characteristic for covalently bonded solids. Parts of this figure, i.e. most of the PME values are taken from previous studies.

## 4. Relating the bond rupture to the field penetration depth

So far, we have presented evidence that the bond rupture differs significantly for different types of solids. Yet, no explanation for these differences has been offered. This is the goal of the present section. To understand why the materials analyzed by APT respond differently under the applied field, we recall the working principle of atom probe tomography, which is based on field evaporation[24]. Field evaporation requires an electric field large enough to break the bonds between a surface atom and its neighbor(s) and can thus remove it from the sample surface. On this surface, there is a density of electrons that can respond to the external field. For conductors like metals, these surface charges shield the electric field. In metals, the charge density at the metallic surface is so large that an external electric field can penetrate only a very short distance into the material. This effect is called "field repulsion" or "field screening"[61]. The situation is quite different for semiconductors since in this case the charge density at the surface is significantly smaller. Hence, the electric field can penetrate over larger distances, i.e. into the solid.

Two different concepts were used to determine the field penetration depth in metals and semiconductors. In metals, the Thomas-Fermi screening is the most common approach to determine the field penetration depth or more precisely the "screening length"[62]. This model calculates how the electric field is screened by electrons in the vicinity of a metallic surface. The two approximations used in this model are *i)* the electrons are considered as an ideal gas and *ii)* the density of electrons is constant. Thus, the screening length is defined by[63]:

$$\delta = \frac{1}{\sqrt{\frac{e^2 D(E_F)}{\varepsilon_0}}}. \qquad [1]$$

where $\varepsilon_0$ is the vacuum permittivity and $D(E_F)$ is the density of states at the Fermi energy $E_F$. Within the Drude model, the conductivity satisfies the Einstein relation $\sigma = e^2 D(E_F) D$, where D is the diffusion constant defined by $D = 1/3\, v_F \lambda$ ($v_F$ is the Fermi velocity and $\lambda$ is the mean free path). Thus, equation [1] can be rewritten as:

$$\delta = \frac{1}{\sqrt{\frac{3\sigma}{v_F l \varepsilon_0}}} \qquad [2]$$

The field screening lengths for various metals and semimetals calculated using this formula are depicted in Figure 4. The values calculated for a series of metals are in the range of 0.5 Å. These values are very short, i.e. significantly shorter than a typical interatomic distance of 2.5 Å[64]. However, for two other solids, which can also be treated with the same equation, i.e. the semimetals Bi and Sb, the screening lengths are about an order of magnitude larger. For these two semimetals, the screening length clearly exceeds the interatomic spacing.

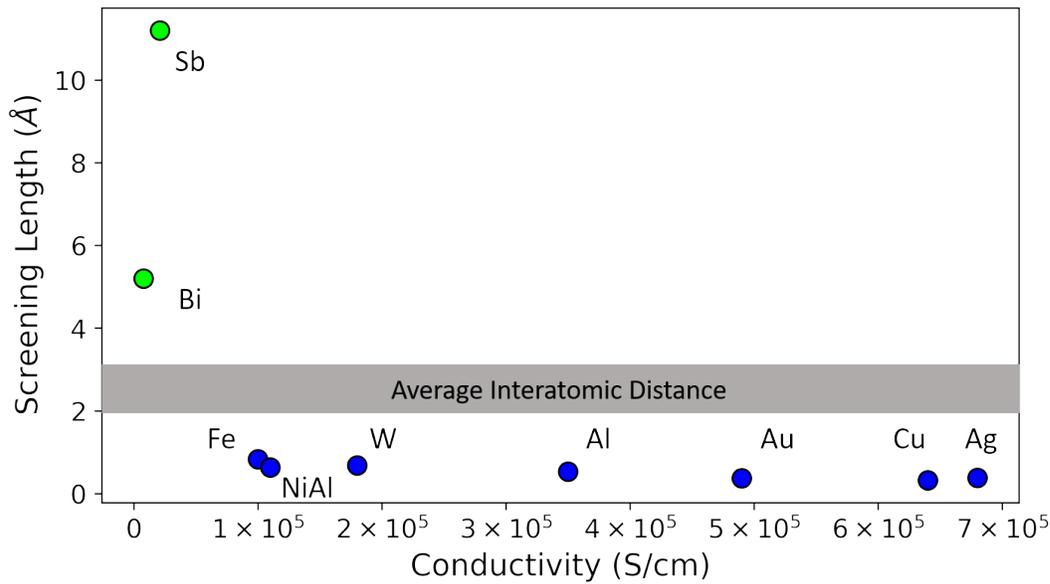

**Figure 4. Screening length for the electrical field in metals as well as in two semimetals (Sb and Bi).** While good metals possess a screening length below 1Å, for the semimetals Bi and Sb the screening length is about one order of magnitude larger.

| Compound | σ (S/cm) | $v_F$ (m/s) | λ (Å) | δ (Å) |
|---|---|---|---|---|
| Al | 3.5×10$^5$ | 2.03×10$^6$ | 1.7×10$^2$ | 0.53 |
| Ag | 6.8×10$^5$ | 1.39×10$^6$ | 2.4×10$^2$ | 0.38 |
| W | 1.8×10$^5$ | 1.5×10$^6$ | 1.9×10$^2$ | 0.68 |
| Fe | 1×10$^5$ | 1.98×10$^6$ | 1.2×10$^2$ | 0.83 |
| Au | 4.9×10$^5$ | 1.38×10$^6$ | 1.65×10$^2$ | 0.37 |
| Cu | 6.4×10$^5$ | 1.1×10$^6$ | 2.15×10$^2$ | 0.32 |
| NiAl | 1.1×10$^5$ | 2.34×10$^6$ | 6.5×10$^2$ | 0.63 |
| Sb | 2.1×10$^4$ | 5.8×10$^5$ | 1.55×10$^5$ | 11.2 |
| Bi | 7.8×10$^3$ | 2.4×10$^4$ | 3×10$^5$ | 5.2 |

**Table 2. Field screening length and conductivity for common metals and semimetals.** The metallic solids are marked blue, whereas the two semimetals, i.e. Sb and Bi are marked green.

For semiconductors, Tsong[65] has developed a scheme to calculate the field penetration depth (screening length) using a spherical geometry of the sample (required for a needle-shaped specimen), where the effective mass of electrons $m_e^*$ and holes $m_h^*$, as well as the static dielectric constant $\varepsilon_r$ are considered. The numerical expression given by Tsong[65] for the field penetration depth $\delta$ is:

$$\delta = \left\{ \frac{\varepsilon_r \varepsilon_0 h^3}{2e^2 [(2\pi)^3 k(m_e^* m_h^*)^{3/2}]^{1/2}} \right\}^{1/2} T^{-1/4} \qquad [3]$$

Density functional theory (DFT) has been used to determine the static dielectric constant as well as the effective mass of electrons and holes for common semiconductors. The values obtained are given in Table 3 together with values for the conductivity $\sigma$ (S/cm).

| Compound | $\varepsilon_r$ | $m_e^*$ | $m_h^*$ | σ (S/cm) | δ (nm) |
|---|---|---|---|---|---|
| SnSe | 60.7 | 0.3*$m_e$ | 0.47*$m_e$ | 0.01 | 6.1 |
| SnS | 47.5 | 0.39*$m_e$ | 1*$m_e$ | 0.01 | 3.7 |
| GeSe | 34.6 | 0.56*$m_e$ | 2.16*$m_e$ | 1.30E-06 | 2.0 |
| Sb$_2$S$_3$ | 72.5 | 01.44*$m_e$ | 0.67*$m_e$ | 1.00E-08 | 3.2 |
| Bi$_2$S$_3$ | 66.9 | 0.39*$m_e$ | 0.69*$m_e$ | 9.40E-07 | 5.0 |
| Sb$_2$Se$_3$ | 79.8 | 1.44*$m_e$ | 0.67*$m_e$ | 4.00E-07 | 3.4 |
| Ge | 34.4 | 0.6*$m_e$ | 0.35*$m_e$ | 0.033 | 3.9 |
| GaN | 9.6 | 0.2*$m_e$ | 1.75*$m_e$ | 1.08E-08 | 1.7 |
| GaSb | 23 | 0.039*$m_e$ | 0.36*$m_e$ | 20 | 8.8 |
| GaAs | 15.4 | 0.067*$m_e$ | 0.58*$m_e$ | 1.00E-08 | 4.9 |
| CdTe | 11.3 | 0.098*$m_e$ | 0.57*$m_e$ | 0.001 | 3.7 |
| ZnTe | 11 | 0.107*$m_e$ | 0.73*$m_e$ | 1E-06 | 3.2 |
| PbO | 18.9 | 1.1*$m_e$ | 3.35*$m_e$ | 4.00E-04 | 1 |
| GeTe | 118.16 | 0.78*$m_e$ | 1.1*$m_e$ | 5000 | 4.3 |
| Sb$_2$Te$_3$ | 287.3 | 0.45*$m_e$ | 0.34*$m_e$ | 2300 | 12.8 |
| Bi$_2$Te$_3$ | 95.2 | 0.32*$m_e$ | 0.44*$m_e$ | 660 | 7.5 |
| SnTe | 150 | 0.12*$m_e$ | 0.13*$m_e$ | 9800 | 21.8 |
| PbTe | 172.2 | 0.18*$m_e$ | 0.28*$m_e$ | 2900 | 15.0 |
| PbSe | 155.6 | 0.12*$m_e$ | 0.17*$m_e$ | 240 | 20.1 |
| PbS | 284.9 | 0.22*$m_e$ | 0.2*$m_e$ | 118 | 20.3 |
| β-As$_2$Te$_3$ | 571 | 0.27*$m_e$ | 0.8*$m_e$ | 650 | 15.8 |
| GeSe (R3m) | 206.1 | 0.19*$m_e$ | 1.25*$m_e$ | 200 | 9.2 |

**Table 3. Static dielectric constant $\varepsilon_r$, effective mass of electrons $m_e^*$ and holes $m_h^*$ as well as field penetration depth ô for common semiconductors.** Covalently bonded solids are marked red, whereas metavalently bonded solids are marked green.

Figure 5 illustrates the field penetration depth versus the conductivity for various covalent (red), metavalent (green), and metallic (blue) solids. The field penetration depth also has a maximum in the transition region between metallic and covalent bonding. This can be explained by the functional dependence of the screening length/penetration depth on the conductivity (for metals) as well as the dielectric constant and the effective masses (for semiconductors). The screening length/penetration depth is closely related to the probability of forming molecular ions. For metallic samples, molecular ions are hardly ever detected. For these solids, the screening length is much shorter than a typical interatomic spacing of about 2.5 Å. This screening length is so efficient that molecular ions are not formed upon laser-assisted field evaporation. For all other solids, including covalent and metavalent solids, the screening length is larger than an interatomic spacing. This is sufficient to create molecular ions, as demonstrated clearly for the two semimetals Sb and Bi. They have a screening length exceeding the interatomic spacing, but only by a factor of about 2 to 4. Yet, this is sufficient to create a large PMI.

Possibly more interesting is the functional dependence of the PME on the electrical conductivity and bonding. Again, we observe particularly high values in the transition region between metals and covalent solids, i.e. the region where metavalent solids are located. This is further evidence that the PME is indeed another bond indicator, besides the characteristic property portfolio discussed in section 3. Yet, it is remarkable that we find solids in the same range of electrical conductivities of about $10^2$ - $10^3$ S/cm, which differ significantly in PME. This can be seen by comparing doped InSb and GaSb with metavalent solids like $Sb_2Te_3$ or GeTe. While InSb and GaSb have modest PME values below 15%, metavalent solids like $Sb_2Te_3$ or GeTe have PME values well above 50%. This finding re-emphasizes two conclusions, different solids that employ different bond types differ in bond rupture and material properties. Yet, the electrical conductivity alone is insufficient to distinguish different types of bonding. In several recent publications, it has been argued that two quantum chemical bonding descriptors can be employed to distinguish different types of bonding, i.e. metallic, covalent, ionic, and metavalent bonding[66]. These two bonding descriptors are derived from the localization and delocalization indices calculated for solids within the QTAIM (quantum theory of atoms in molecules)[67]. From these quantities, we determine the electron transfer between adjacent atoms and the number of electrons shared between them. The resulting map is shown in Figure 6. The colors of the different solids characterize the different properties. These different material properties are located in different regions of the map, in line with the argument that there is a close link between these quantum-chemical bonding descriptors and certain material properties, as discussed in detail in ref.[60]. In Figure 6, a z-axis has been added which characterizes the PME. Systematic changes of the PME are discernible in the map. This implies that there is indeed a close relationship between bond rupture in atom probe tomography and different types of chemical bonding.

There is another fascinating aspect which can be derived from comparing figures 5 and 6. The unique bond rupture which characterizes metavalent solids is located in a well-defined range between approximately $5 \times 10^2$ S/cm and $5 \times 10^4$ S/cm. This is the range of (room temperature) conductivities, where a transition between metals and insulators occurs in most solids. It is also the conductivity range, where Mooij has found a change in the sign of the temperature coefficient of the electric conductivity in many metals[68]. Please note that we now use the term metals and insulators to describe the 0K limit of the electrical conductivity and not the type of chemical bonding. Doped semiconductors, for example, can turn metallic in terms of the electrical conductivity if doped sufficiently high to become degenerate semiconductors, but do not change their bonding type, i.e. change their atomic arrangement. This metal to insulator transition (MIT) upon doping has been well-studied for many semiconductors such as Si and GaAs[69]. Our data show that for GaAs this transition is not accompanied by a significant change of the PME, while this happens for metavalent solids. This indicates that in metavalent solids an MIT is realized which differs significantly from electronic MITs attributed to either the Mott-type (correlation)[70] or the Anderson-type (disorder)[71] MIT in doped semiconductors. This is another interesting research opportunity for atom probe tomography.

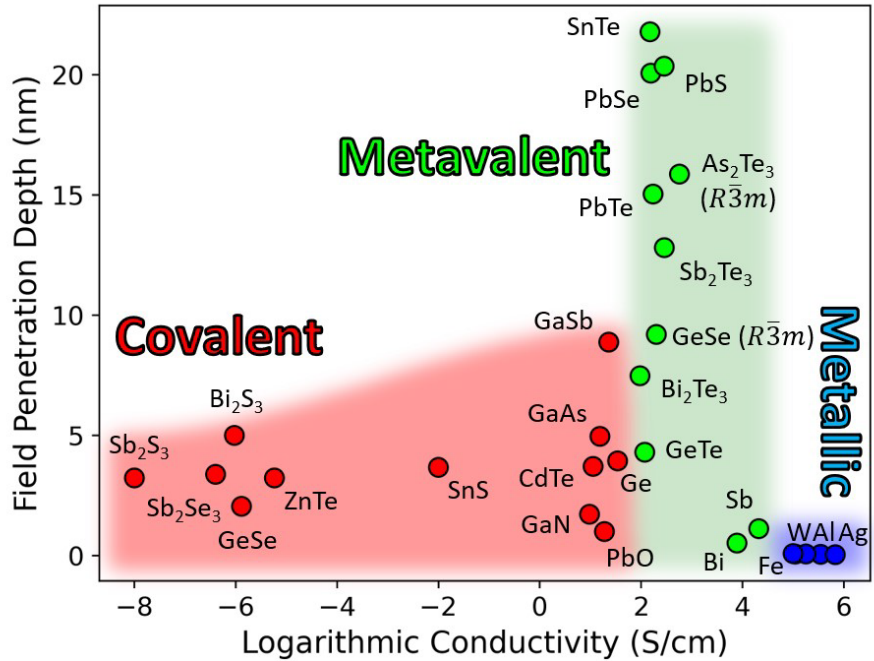

**Figure 5.** Field penetration depth versus conductivity for various covalent (red), metavalent (green) and metallic (blue) solids.

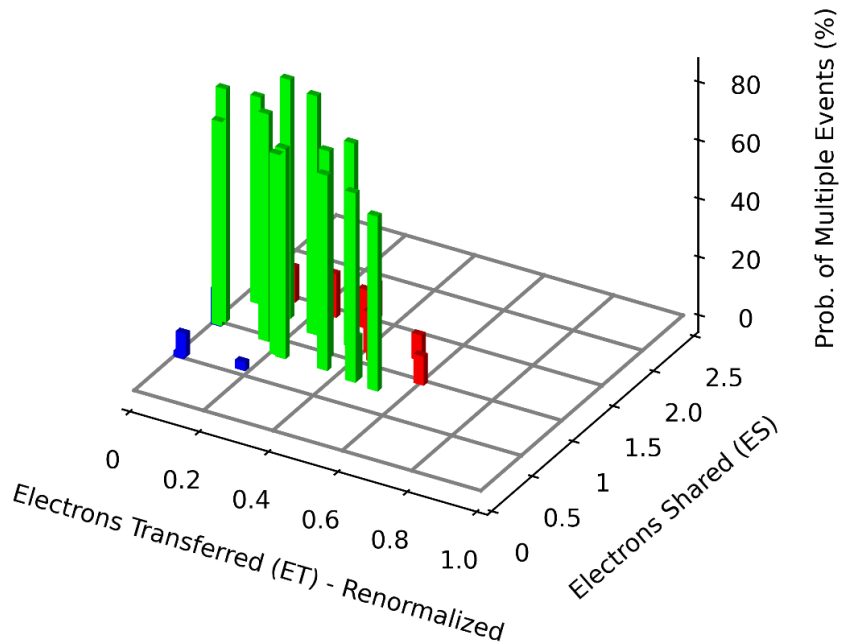

**Figure 6. PME as a function of bonding mechanism.** High PME values are a characteristic of metavalent solids, while metallic and covalent solids possess much smaller PME values. This figure also shows that the two quantum-chemical bonding descriptors are the best property predictors for the unusual bond rupture which characterizes metavalent solids.

## 5.     Utilizing the bond rupture as a local probe of chemical bonding

In the last section, a close relationship between different types of bonding and the bond rupture in atom probe tomography has been established. One can now contemplate, how this characteristic of APT can be utilized to understand and tailor functional materials. At present, there is only a limited number of examples which demonstrate how APT can be employed to tailor functional materials based upon an in-depth understanding of bond rupture. This can be partly attributed to the novelty of the conclusions presented above. Yet, it seems easy to sketch the promise this approach offers. One key advantage of atom probe tomography is the high spatial resolution since APT offers high, i.e. atomic resolution in three dimensions. One can now ponder which spatial resolution can be reached to locally determine the bond rupture and hence type of bonding employed. A determination of the PME and PMI requires reasonable statistics. Hence, ideally, 1000 successful laser pulses should be analyzed to determine those two quantities. With a detection rate of 50%, this corresponds to 2000 laser pulses and corresponds to a cube of 40 Å × 40 Å × 40 Å. This would indeed be very interesting for various material design approaches.

Recently, we have demonstrated the informative value of such data[72]. In an attempt to unravel the impact of grain boundaries on charge transport, single grain boundaries in Ag-doped PbTe were characterized by APT and resistivity measurements. It was shown that the resistivity of the grain boundary was closely related to the difference in orientation of the two adjacent grains across this boundary. Small-angle grain boundaries of Ag-doped PbTe showed a much lower resistance than high-angle grain boundaries[72]. Subsequently, these changes were related to significant changes in bonding in the vicinity of the grain boundary. While a high PME was observed within the grain (Figure 7a), indicative of metavalent bonding within the grain of PbTe, a much lower PME value was observed in a region adjacent to the high-angle grain boundary (Figure 7b). This is indicative of a significant change in bonding and properties and can help to understand the pronounced impact of such grain boundaries. For low-angle grain boundaries instead, a much smaller change in bond rupture, i.e. a much smaller change in PME was found (Figure 7c,d). This indicates that in this case, a much less extended region with pronounced changes in bonding occurs. This already demonstrates the kind of novel insights that local measurements of bond type offer.

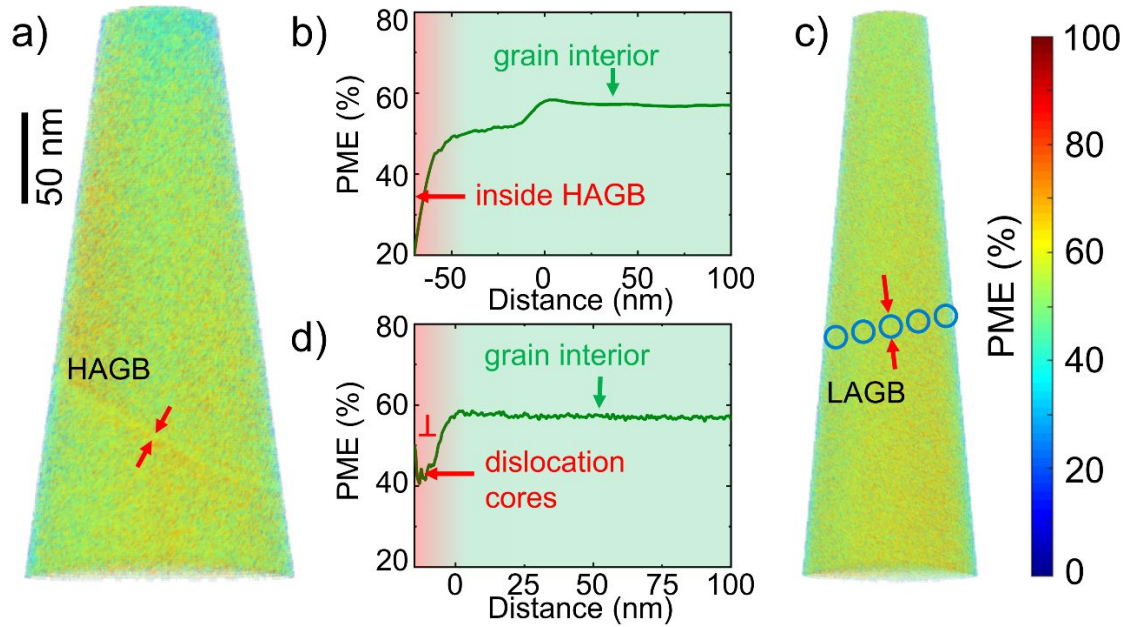

**Figure 7. Local changes in bond rupture showing the impact of grain boundaries (GB) in a Ag-doped PbTe compound.** (a) 3D PME map including a high-angle grain boundary. (b) PME proximity histogram calculated from the grain interior into the GB plane, where a significant drop in PME can be found inside the high-angle GB. (c). 3D PME map including a low-angle GB which is formed by dislocation arrays as indicated by blue circles. (d) PME proximity histogram showing the decrease in PME inside the dislocation cores, while the regions between dislocations cores maintain high PME values. Figures are adapted from Ref.[72] with permission.

Many more examples can be suggested. In phase change materials, for example, the process of incubation, i.e. the formation of subcritical crystalline nuclei and their growth into stable nuclei is of paramount importance. As amorphous phase change materials have a much lower PME than their crystalline counterpart, the formation of crystalline nuclei could be detected with APT. Using APT, it would even be possible to verify if such crystalline nuclei differ in stoichiometry from their amorphous surroundings.

Another exciting area is transitions between the metallic and the insulating state[73,71,74] As shown in Figure 3 and Figure 6, a striking change of bond rupture is identified upon the transition between metallic and non-metallic states. This is intriguing since one can ponder how the transition from the metallic to the insulating state will be reflected in atom probe measurements.

**Conclusions**

In this review, the potential of atom probe tomography to locally probe chemical bonds is presented and discussed. Two processes are shown to characterize the bond rupture in laser-assisted field emission. These are the probability of molecular ions (PMI), i.e. the probability that molecular ions are evaporated instead of single (atomic) ions, and the probability of multiple events (PME), i.e. the correlated field-evaporation of more than a single fragment upon laser- or voltage pulse excitation. Here we demonstrate that one can clearly distinguish solids with metallic, covalent, and metavalent bonds based on their bond rupture, i.e. their PME and PMI values. These differences are largely attributed to differences in the field penetration depth. These findings open new avenues in understanding and designing advanced materials, since they allow quantification of bonds in solids on a nanometer scale. This is shown for the metavalent solid PbTe, where large angle grain boundaries have a pronounced impact on charge carrier mobility. This finding can be attributed to the collapse of metavalent bonding at the grain boundary. Nanometer-sized crystalline grains of phase change materials in an amorphous matrix present a second interesting class of solids for such studies. These examples even seem to justify calling the present approach *bonding probe tomography* (BPT).

**Materials and Methods**

   a) **Crystal preparation**

Most crystalline samples were synthesized from the elements in a vacuum-sealed quartz ampoule. The amorphous samples were deposited on Si substrates by sputtering employing alloy target of 99.99% purity. Annealing these amorphous specimens enabled the preparation of crystalline samples, too. The stoichiometry of the resulting samples was obtained from energy dispersive spectroscopy (EDS).

   b) **Experimental investigations**

APT analyses were conducted using a CAMECA LEAP-5000 XS (local electrode atom probe). For APT measurements, the specimen was maintained at 50 K and laser pulses (wavelength 355 nm) of 10 pJ energy were used for field evaporation. Detection rate of 1 ion per 500 pulses was chosen to obtain a 250 kHz pulse repetition rate. Moreover, the APT needle-shaped samples were prepared by standard lift-out procedure, using a dual-beam focus ion beam (FEI Helios Nanolab 650). These samples, with a top radius smaller than 80 nm, were mounted on flat top Si microtips.

   c) **Theoretical investigations**

The first principle calculations were based on the Vienna Ab initio Simulation Package (VASP)[75]. The interaction between ions and valence electrons was described by Projected Augmented Wave (PAW)[76], and the exchange-correlation interaction was described by the Perdew-Burke-Ernzerhof generalized gradient approximation (PBE-GGA)[77,78]. The cut-off energy was set as 400 eV, and the convergence criteria for self-consistent electronic energy and residual force were respectively assumed to be $10^{-6}$ eV/atom and 0.01 eV/Å. The k-points were set up 4 × 4 × 4 based on Monkhorst-Pack grids. The VASPKIT was employed to generate the Brillouin zone pathway and extract the band gap and effective mass of all structures[79].

The static dielectric constants have been computed using Density Functional Perturbation Theory including local field effects[80].

**Author Contributions**

OCM and MW have designed and developd this study. YY, JK, TG, C-FS, SH and CZ have obtained the experimental and theoretical data presented here. OCM and MW wrote the present publication with input from all coauthors.

**Conflict of Interest**

The authors declare no conflict of interests.

**Acknowledgements**

The authors gratefully acknowledge the computational resources by the JARA-HPC from RWTH Aachen University under project JARA0236 and the computing time provided to them at the NHR Center NHR4CES at RWTH Aachen University (project number p0020357). This work was supported in part by the Deutsche Forschungsgemeinschaft (SFB 917) and in part by the Federal Ministry of Education and Research (BMBF, Germany) in the project NEUROTEC II (16ME0398K). TG acknowledges the Alexander von Humboldt Foundation for the postdoctoral fellowship received which allowed him to conduct part of this research.

**Bibliography:**

1. Kelly, T. F. & Miller, M. K. Atom probe tomography. *Rev. Sci. Instrum.* **78**, 31101 (2007).

2. Seidman, D. N. Three-dimensional atom-probe tomography: Advances and applications. *Annu. Rev. Mater. Res.* **37**, 127–158 (2007).

3. McCarroll, I. E., Bagot, P. A. J., Devaraj, A., Perea, D. E. & Cairney, J. M. New frontiers in atom probe tomography: a review of research enabled by cryo and/or vacuum transfer systems. *Mater. today. Adv.* **7**, (2020).

4. Kelly, T. F. & Larson, D. J. Atom probe tomography 2012. *Annu. Rev. Mater. Res.* **42**, 1–31 (2012).

5. Kelly, T. E. *et al.* Atom probe tomography of electronic materials. *Annu. Rev. Mater. Res.* **37**, 681–727 (2007).

6. Choi, P.-P. *et al.* Atom Probe Tomography of Compound Semiconductors for Photovoltaic and Light-Emitting Device Applications. *Micros. Today* **20**, 18–24 (2012).

7. Thompson, K., Flaitz, P. L., Ronsheim, P., Larson, D. J. & Kelly, T. F. Imaging of arsenic Cottrell atmospheres around silicon defects by three-dimensional atom probe tomography. *Science (80-. ).* **317**, 1370–1374 (2007).

8. Raghuwanshi, M., Wuerz, R. & Cojocaru-Mirédin, O. Interconnection between Trait, Structure, and Composition of Grain Boundaries in Cu(In,Ga)Se2 Thin-Film Solar Cells. *Adv. Funct. Mater.* **30**, 1–

9 (2020).

9. Cojocaru-Mirďin, O., Mangelinck, D. & Blavette, D. Nucleation of boron clusters in implanted silicon. *J. Appl. Phys.* **106**, 1–7 (2009).

10. Cojocaru-Mirédin, O., Cristiano, F., Fazzini, P.-F., Mangelinck, D. & Blavette, D. Extended defects and precipitation in heavily B-doped silicon. *Thin Solid Films* **534**, 62–66 (2013).

11. Philippe, T., Cojocaru-MirÉdin, O., Duguay, S. & Blavette, D. Clustering and nearest neighbour distances in atom probe tomography: The influence of the interfaces. *J. Microsc.* **239**, 72–77 (2010).

12. Tang, F. *et al.* Indium clustering in a -plane InGaN quantum wells as evidenced by atom probe tomography. *Appl. Phys. Lett.* **106**, (2015).

13. Zhao, H. *et al.* Segregation assisted grain boundary precipitation in a model Al-Zn-Mg-Cu alloy. *Acta Mater.* **156**, 318–329 (2018).

14. Devaraj, A. *et al.* Visualizing the Nanoscale Oxygen and Cation Transport Mechanisms during the Early Stages of Oxidation of Fe–Cr–Ni Alloy Using In Situ Atom Probe Tomography. *Adv. Mater. Interfaces* **9**, 2200134 (2022).

15. Sun, Z. *et al.* Dopant Diffusion and Activation in Silicon Nanowires Fabricated by ex Situ Doping: A Correlative Study via Atom-Probe Tomography and Scanning Tunneling Spectroscopy. *Nano Lett.* **16**, 4490–4500 (2016).

16. Meng, Y., Kim, H., Isheim, D., Seidman, D. N. & Zuo, J.-M. Atom-Probe Tomographic Study of Interfacial Intermixing and Segregation in InAs/GaSb Superlattices. *Microsc. Microanal.* **19**, 958–959 (2013).

17. Rajeev, A. *et al.* Interfacial Mixing Analysis for Strained Layer Superlattices by Atom Probe Tomography. *Crystals* vol. 8 at https://doi.org/10.3390/cryst8110437 (2018).

18. Cojocaru-Mirédin, O., Hollermann, H., Mio, A. M., Wang, A. Y. T. & Wuttig, M. Role of grain boundaries in Ge-Sb-Te based chalcogenide superlattices. *J. Phys. Condens. Matter* **31**, ab078b (2019).

19. Soni, P. *et al.* Role of elemental intermixing at the In 2 S 3 /CIGSe heterojunction deposited using reactive RF magnetron sputtering. *Sol. Energy Mater. Sol. Cells* **195**, 367–375 (2019).

20. Cojocaru-Mirédin, O. *et al.* Snowplow effect and reactive diffusion in the Pt doped Ni-Si system. *Scr. Mater.* **57**, 373–376 (2007).

21. Liebscher, C. H. *et al.* Strain-Induced Asymmetric Line Segregation at Faceted Si Grain Boundaries. *Phys. Rev. Lett.* **121**, 1–5 (2018).

22. Kuzmina, M., Herbig, M., Ponge, D., Sandlöbes, S. & Raabe, D. Linear complexions: Confined chemical and structural states at dislocations. *Science (80-. ).* **349**, 1080–1083 (2015).

23. Shu, R. *et al.* Solid-State Janus Nanoprecipitation Enables Amorphous-Like Heat Conduction in Crystalline Mg3Sb2-Based Thermoelectric Materials. *Adv. Sci.* **9**, 2202594 (2022).

24. Gault, B. *et al.* Atom probe tomography. *Nat. Rev. Methods Prim.* **1**, (2021).

25. Xie, K. Y. *et al.* Breaking the icosahedra in boron carbide. *Proc. Natl. Acad. Sci. U. S. A.* **113**, 12012–12016 (2016).

26. Rusitzka, K. A. K. *et al.* A near atomic-scale view at the composition of amyloid-beta fibrils by atom probe tomography. *Sci. Rep.* **8**, 17615 (2018).

27. Zhu, M. *et al.* Unique Bond Breaking in Crystalline Phase Change Materials and the Quest for Metavalent Bonding. *Adv. Mater.* **30**, (2018).

28. Cheng, Y. *et al.* Understanding the Structure and Properties of Sesqui-Chalcogenides (i.e., V2VI3 or Pn2Ch3 (Pn = Pnictogen, Ch = Chalcogen) Compounds) from a Bonding Perspective. *Adv. Mater.* (2019) doi:10.1002/adma.201904316.

29. Yu, Y., Cagnoni, M., Cojocaru-Mirédin, O. & Wuttig, M. Chalcogenide Thermoelectrics Empowered by an Unconventional Bonding Mechanism. *Adv. Funct. Mater.* **30**, (2020).

30. Gault, B. *Chapter 1: Introduction. Atom Probe Micrscopy* (2004). doi:10.1007/978-1-4614-3436-8.

31. Blum, I. *et al.* Dissociation Dynamics of Molecular Ions in High dc Electric Field. *J. Phys. Chem. A* **120**, 3654–3662 (2016).

32. Johansen, M. & Liu, F. Best Practices for Analysis of Carbon Fibers by Atom Probe Tomography. *Microsc. Microanal.* **28**, 1092–1101 (2022).

33. Qiu, R. *et al.* Atom probe tomography investigation of 3D nanoscale compositional variations in CVD TiAlN nanolamella coatings. *Surf. Coatings Technol.* **426**, 127741 (2021).

34. Aboulfadl, H. *et al.* Microstructural Characterization of Sulfurization Effects in Cu(In,Ga)Se2 Thin Film Solar Cells. *Microsc. Microanal.* **25**, 532–538 (2019).

35. Cojocaru-Mirédin, O. *et al.* Characterization of Grain Boundaries in Cu(In,Ga)Se2 Films Using Atom-Probe Tomography. *IEEE J. Photovoltaics* **1**, 207–212 (2011).

36. Douglas, J. O. *et al.* Atom Probe Tomography for Isotopic Analysis: Development of the 34S/32S System in Sulfides. *Microsc. Microanal.* **28**, 1127–1140 (2022).

37. Beckey, H. D. *Field Ionisation Mass spectrometry*. (Pergamon Press, 1971).

38. Saxey, D. W. Correlated ion analysis and the interpretation of atom probe mass spectra. *Ultramicroscopy* **111**, 473–479 (2011).

39. Russo, E. Di *et al.* Compositional accuracy of atom probe tomography measurements in GaN: Impact of experimental parameters and multiple evaporation events. *Ultramicroscopy* **187**, 126–134 (2018).

40. Pedrazzini, S. *et al.* Nanoscale Stoichiometric Analysis of a High-Temperature Superconductor by Atom Probe Tomography. *Microsc. Microanal. Off. J. Microsc. Soc. Am. Microbeam Anal. Soc. Microsc. Soc. Canada* **23**, 414–424 (2017).

41. Tang, F., Gault, B., Ringer, S. P. & Cairney, J. M. Optimization of pulsed laser atom probe (PLAP) for the analysis of nanocomposite Ti-Si-N films. *Ultramicroscopy* **110**, 836–843 (2010).

42. Meisenkothen, F., Samarov, D. V, Kalish, I. & Steel, E. B. Exploring the accuracy of isotopic analyses in atom probe mass spectrometry. *Ultramicroscopy* **216**, 113018 (2020).

43. Müller, M., Saxey, D. W., Smith, G. D. W. & Gault, B. Some aspects of the field evaporation behaviour of GaSb. *Ultramicroscopy* **111**, 487–492 (2011).

44. Müller, M., Smith, G. D. W., Gault, B. & Grovenor, C. R. M. Compositional nonuniformities in pulsed laser atom probe tomography analysis of compound semiconductors. *J. Appl. Phys.* **111**, (2012).

45. Yao, L., Gault, B., Cairney, J. M. & Ringer, S. P. On the multiplicity of field evaporation events in atom probe: A new dimension to the analysis of mass spectra. *Philos. Mag. Lett.* **90**, 121–129 (2010).

46. Thuvander, M. *et al.* Quantitative atom probe analysis of carbides. *Ultramicroscopy* **111**, 604–608 (2011).

47. Licata, O. G., Broderick, S. R. & Mazumder, B. Correlation of Multiplicity and Chemistry in Al (x) Ga(1-x)N Heterostructure via Atom Probe Tomography. *Microsc. Microanal. Off. J. Microsc. Soc.*

*Am. Microbeam Anal. Soc. Microsc. Soc. Canada* **26**, 95–101 (2020).

48. Gault, B. *et al.* Behavior of molecules and molecular ions near a field emitter. *New J. Phys.* **18**, (2016).

49. Soni, P., Cojocaru-Miredin, O. & Raabe, D. Interface engineering and nanoscale characterization of Zn(S,O) alternative buffer layer for CIGS thin film solar cells. *2015 IEEE 42nd Photovolt. Spec. Conf. PVSC 2015* (2015) doi:10.1109/PVSC.2015.7355889.

50. Kitaguchi, H. S. *et al.* An atom probe tomography study of the oxide–metal interface of an oxide intrusion ahead of a crack in a polycrystalline Ni-based superalloy. *Scr. Mater.* **97**, 41–44 (2015).

51. Marquis, E. A., Yahya, N. A., Larson, D. J., Miller, M. K. & Todd, R. I. Probing the improbable: imaging C atoms in alumina. *Mater. Today* **13**, 34–36 (2010).

52. Kinno, T. *et al.* Influence of multi-hit capability on quantitative measurement of NiPtSi thin film with laser-assisted atom probe tomography. *Appl. Surf. Sci.* **259**, 726–730 (2012).

53. Zhu, M. *et al.* Unique Bond Breaking in Crystalline Phase Change Materials and the Quest for Metavalent Bonding. *Adv. Mater.* **30**, 1–9 (2018).

54. Hughes, E. T., Haidet, B. B., Bonef, B., Cai, W. & Mukherjee, K. Pipe-diffusion-enriched dislocations and interfaces in SnSe/PbSe heterostructures. *Phys. Rev. Mater.* **5**, 73402 (2021).

55. Maier, S. *et al.* Discovering Electron-Transfer-Driven Changes in Chemical Bonding in Lead Chalcogenides (PbX, where X = Te, Se, S, O). *Adv. Mater.* **32**, 1–11 (2020).

56. Pries, J., Cojocaru-Mirédin, O. & Wuttig, M. Phase-change materials: Empowered by an unconventional bonding mechanism. *MRS Bull.* **44**, 699–704 (2019).

57. Wuttig, M., Deringer, V. L., Gonze, X., Bichara, C. & Raty, J. Y. J.-Y. SI: Incipient metals: Functional materials with a unique bonding mechanism. *Adv. Mater.* **30**, 1–6 (2018).

58. Raty, J.-Y., Bichara, C., Schön, C.-F., Gatti, C. & Wuttig, M. Tailoring chemical bonds to design unconventional glasses. *Proc. Natl. Acad. Sci.* **121**, e2316498121 (2024).

59. Wuttig, M. & Yamada, N. Phase-change materials for rewriteable data storage. *Nat. Mater.* **6**, 824–832 (2007).

60. Schön, C.-F. *et al.* Classification of properties and their relation to chemical bonding: Essential steps toward the inverse design of functional materials. *Sci. Adv.* **8**, eade0828 (2024).

61. Miller, M. & Forbes, R. *Atom Probe Tomography: The Local Electrode Atom Probe*. *Atom-Probe Tomography: The Local Electrode Atom Probe* (2014). doi:10.1007/978-1-4899-7430-3.

62. Mönch, W. *Semiconductor surfaces and interfaces*. vol. 26 (Springer Science & Business Media, 2013).

63. Ibach, H. & Lüth, H. *Festkörperphysik: Einführung in die Grundlagen*. (Springer Berlin Heidelberg, 2009).

64. Daams, J. L., Rodgers, J. R. & Villars, P. Typical interatomic distances: metals and alloys. *Int. Tables Crystallogr.* **C**, 774 (2006).

65. Tsong, T. T. Field penetration and band bending for semiconductor of simple geometries in high electric fields. *Surf. Sci.* **85**, 1–18 (1979).

66. Raty, J. Y. *et al.* A Quantum-Mechanical Map for Bonding and Properties in Solids. *Adv. Mater.* **31**, 1–6 (2019).

67. Bader, R. F. W. *Atoms in Molecules: A Quantum Theory*. (Press: Oxford, U.K., 1990).

68. Mooij, J. H. Electrical conduction in concentrated disordered transition metal alloys. *Phys. status solidi* **17**, 521–530 (1973).

69. Rosenbaum, T. F., Andres, K., Thomas, G. A. & Bhatt, R. N. Sharp Metal-Insulator Transition in a Random Solid. *Phys. Rev. Lett.* **45**, 1723–1726 (1980).

70. Mott, N. F. The transition to the metallic state. *Philos. Mag. A J. Theor. Exp. Appl. Phys.* **6**, 287–309 (1961).

71. Anderson, P. W. Absence of Diffusion in Certain Random Lattices. *Phys. Rev.* **109**, 1492–1505 (1958).

72. Wu, R. *et al.* Strong charge carrier scattering at grain boundaries of PbTe caused by the collapse of metavalent bonding. *Nat. Commun.* **14**, 719 (2023).

73. Siegrist, T. *et al.* Disorder-induced localization in crystalline phase-change materials. *Nat. Mater.* **10**, 202–208 (2011).

74. Fratini, S., Ciuchi, S., Dobrosavljević, V. & Rademaker, L. Universal Scaling near Band-Tuned Metal-Insulator Phase Transitions. *Phys. Rev. Lett.* **131**, 196303 (2023).

75. Kresse, G. & Furthmüller, J. Efficient iterative schemes for ab initio total-energy calculations using a plane-wave basis set. *Phys. Rev. B - Condens. Matter Mater. Phys.* **54**, 11169–11186 (1996).

76. Blöchl, P. E. Projector augmented-wave method. *Phys. Rev. B* **50**, 17953–17979 (1994).

77. Perdew, J. P., Burke, K. & Ernzerhof, M. Generalized gradient approximation made simple. *Phys. Rev. Lett.* **77**, 3865–3868 (1996).

78. Perdew, J. P., Burke, K. & Wang, Y. Generalized gradient approximation for the exchange-correlation hole of a many-electron system. *Phys. Rev. B* **54**, 16533–16539 (1996).

79. Wang, V., Xu, N., Liu, J., Tang, G. & Geng, W. *VASPKIT: A User-friendly Interface Facilitating High-throughput Computing and Analysis Using VASP Code*. (2019).

80. Baroni, S. & Resta, R. Ab initio calculation of the macroscopic dielectric constant in silicon. *Phys. Rev. B* **33**, 7017–7021 (1986).